%% file: 00paper.tex
\documentclass[sigconf]{acmart}

\usepackage{amsmath}
\usepackage{listings}
\usepackage{graphicx}
\usepackage{verbatim}
\usepackage{booktabs}
\usepackage{balance}

\copyrightyear{2020} 
\acmYear{2020} 
\setcopyright{acmcopyright}
\acmConference[ICTIR '20]{The 2020 ACM SIGIR International Conference on the Theory of Information Retrieval}{September 14--17, 2020}{Virtual Event, Norway}
\acmBooktitle{The 2020 ACM SIGIR International Conference on the Theory of Information Retrieval (ICTIR '20), September 14--17, 2020, Virtual Event, Norway}
\acmPrice{15.00}
\acmDOI{10.1145/3409256.3409836}
\acmISBN{978-1-4503-8067-6/20/09}

\begin{document}

\fancyhead{}

\title{Sanitizing Synthetic Training Data Generation for Question Answering over Knowledge Graphs}

\author{Trond Linjordet}
\affiliation{%
  \institution{University of Stavanger}
  \city{Stavanger}
  \state{Norway}
}
\email{trond.linjordet@uis.no}

\author{Krisztian Balog}
\affiliation{%
  \institution{University of Stavanger}
  \city{Stavanger}
  \state{Norway}
}
\email{krisztian.balog@uis.no}

\renewcommand{\shortauthors}{Linjordet and Balog}

\begin{abstract}
  Synthetic data generation is important to training and evaluating neural models for question answering over knowledge graphs. 
  The quality of the data and the partitioning of the datasets into training, validation and test splits impact the performance of the models trained on this data. 
  If the synthetic data generation depends on templates, as is the predominant approach for this task, there may be a leakage of information via a shared basis of templates across data splits if the partitioning is not performed hygienically. 
  This paper investigates the extent of such information leakage across data splits, and the ability of trained models to generalize to test data when the leakage is controlled.
    We find that information leakage indeed occurs and that it affects performance. 
  At the same time, the trained models do generalize to test data under the sanitized partitioning presented here. 
  Importantly, these findings extend beyond the particular flavor of question answering task we studied and raise a series of difficult questions around template-based synthetic data generation that will necessitate additional research.
\end{abstract}



\maketitle

\input{ictir2020-synthetic-01}
\input{ictir2020-synthetic-02}
\input{ictir2020-synthetic-03}
\input{ictir2020-synthetic-04}
\input{ictir2020-synthetic-05}
\input{ictir2020-synthetic-06}

\bibliographystyle{ACM-Reference-Format}
\balance
\bibliography{ictir2020-synthetic}
\end{document}

%% file: ictir2020-synthetic-01.tex
\section{Introduction}

\begin{figure}[t]
    \centering
    \vspace*{0.5\baselineskip}
	\includegraphics[width=0.4\textwidth]{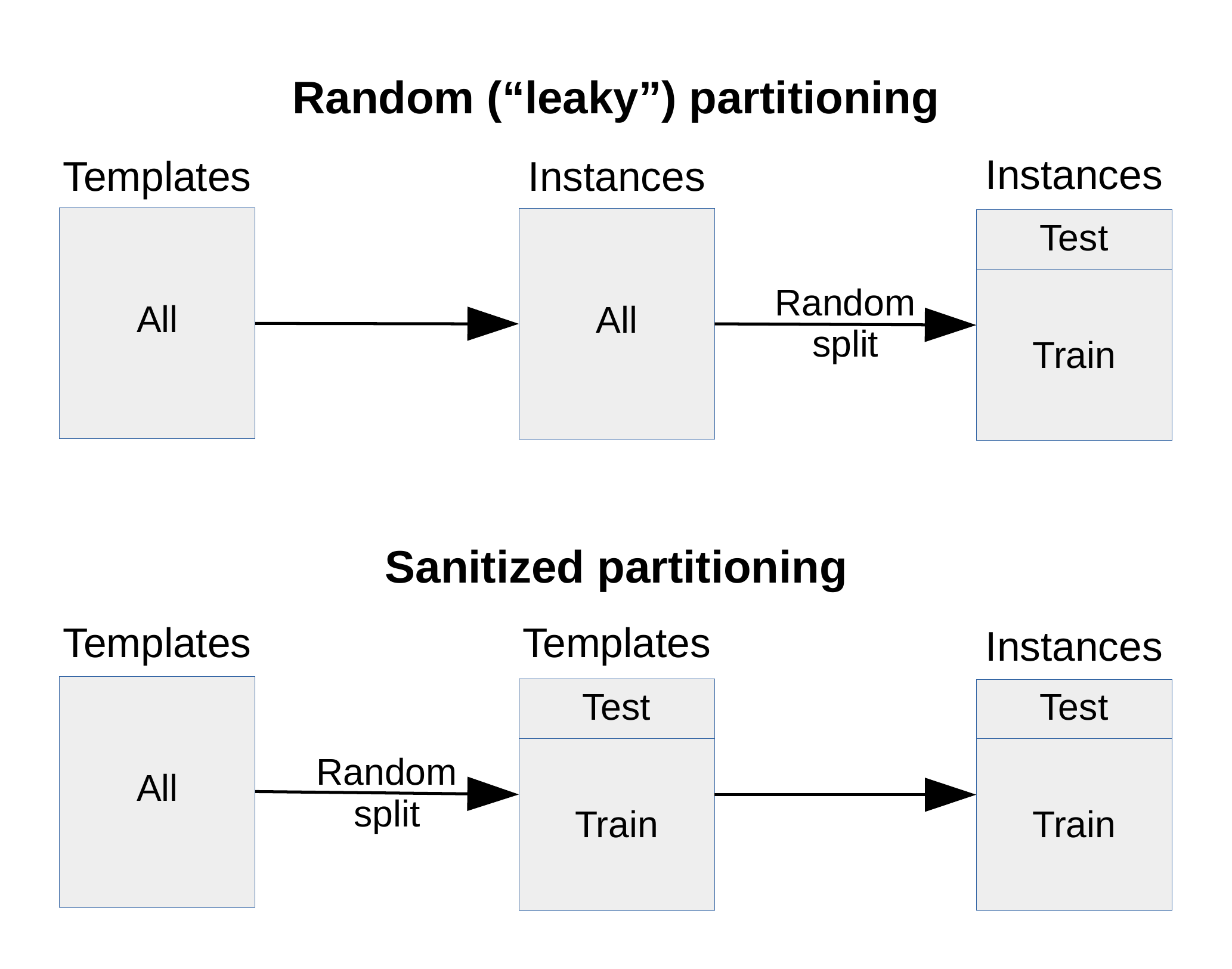}
    \vspace*{0.25\baselineskip}
	\caption{Illustration of leaky and sanitized partitioning for template-based synthetic data generation.}
	\label{fig:overview}
    \vspace*{0.25\baselineskip}
\end{figure}

Synthetic data generation can benefit neural models by producing adequate volumes of training data. Knowledge graph question answering (KGQA)---the problem of mapping natural language questions to SPARQL queries---is a task where deep neural models have recently been introduced. Neural models for KGQA by their high-volume data requirements bring about the need for synthetic data generation. However, unlike for other tasks like ad hoc document retrieval~\citep{Azzopardi:2007:BSQ}, query clarification terms~\citep{Ding:2018:GSD}, or query auto-completion~\citep{Krishnan:2020:GSQ}, synthetic data generation for KGQA has so far been developed in a template-based manner. This raises a number of interesting methodological questions.

In particular, we consider a hypothesis that training models on template-based synthetic data instances may result in \emph{information leakage} if the partitioning of synthetic data into training, validation, and test splits is not done carefully. If the training and test splits are randomly partitioned without regard for the underlying templates, it is possible that a significant portion of the performance seen in trained models is not coming from correct generalizations. Instead, some portion of the observed performance may come from memorizing the underlying patterns of the finite set of templates, which are common across the training, validation, and testing splits. This \emph{leaky} partitioning condition is illustrated in the top part of Fig.~\ref{fig:overview}. To explore the hypothesis, we devised an alternative, \emph{sanitized} partitioning scheme, illustrated in the bottom part of Fig.~\ref{fig:overview}. 

As a guide to intuition, we can imagine that the KGQA models trained on template-based instances will ``see'' through the instance to the underlying template, which is therefore considered \emph{seen} with respect to the trained model. The question of the trained models' ability to generalize can then be cast as a question of how the trained models perform on instances generated from \emph{unseen} templates. 

We address three research questions:
\begin{itemize}
	\item \textbf{RQ1:} Is the performance of trained neural KGQA models affected by whether testing templates are \emph{seen} or \emph{unseen}? 
	\item \textbf{RQ2:} Is the ability to generalize to instances based on \emph{unseen} templates affected by the volume of training data used?
	\item \textbf{RQ3:} Can the proportion of \emph{unseen} templates to \emph{seen} templates affect the trained models' ability to generalize?
\end{itemize}

\noindent
Specifically, we look at complex KGQA, which is a variant of KGQA where the formal query represents a multi-relation subgraph on the knowledge graph (KG). 
We investigate the properties of synthetic data in the context of neural network models, using the largest KGQA dataset that exists to date, DBpedia Neural Question Answering (DBNQA)~\citep{Hartmann:2018:GLD}. 
We empirically compare three neural machine translation (NMT) architectures that represent a specific family of neural network architectures, recurrent neural networks (RNNs), which were shown to be effective on this task~\citep{Soru:2017:SFL, Soru:2018:NMT, Chakraborty:2019:INN}. 

In the \emph{leaky} partitioning, instances are randomly assigned to splits without regard for underlying templates. 
This \emph{leaky} partitioning is both convenient and provides the models with the maximum volume and variety of training instances. In the \emph{sanitized} partitioning, templates are partitioned into train and test splits, and instances are then partitioned so that test instances will only be those generated from \emph{unseen} templates. If the synthetic data is not generated in a \emph{sanitized} manner initially, the \emph{sanitized} partitioning requires additional processing to achieve: first templates must be partitioned into test and training splits; then the generated instances must be matched with the templates they were generated from; and finally, the instances must be allocated to test and training splits, accordingly. Nevertheless, this approach helps minimize information leakage when testing model performance. 


Empirically, we observe the expected loss of performance on sanitized test splits compared to leaky validation splits (RQ1). 

When adjusting the volume of sanitized training data, we see small but consistent performance increase in response to increased training data, even with respect to instances from unseen templates (RQ2). 
Likewise, when the proportion of templates assigned to the training split is increased, the performance on sanitized test splits is low, but does show improvement (RQ3).
These results indicate that while information leakage and memorizing the patterns of seen templates account for a lot of the performance observed under the leaky partitioning, some generalization does happen from instances of seen templates to instances of unseen templates.    

To summarize, the most important contribution of this study is the identification of the problem of information leakage in template-based synthetic generation approaches.  Using a large KGQA dataset, we show that that random partitioning as in~\citep{Yin:2019:NMT} is indeed leaky, giving a misleading impression of the performance of the trained models.  The significance of our finding, however, extends beyond KGQA, as it applies to any template-base data generation approach, and raises a set of interesting questions around training models with synthetic data using fair conditions.  We present a novel dataset partitioning scheme that provides a facility to quantify the generalized learning achieved by models trained on template-generated synthetic data. 

%% file: ictir2020-synthetic-02.tex
\section{Background}
\label{sec:background}

To satisfy the need for large volume datasets to train deep learning models, various approaches have been explored to enhance the collection of real data points, such as data augmentation~\citep{Dao:2019:KTM, Shorten:2019:SID} and synthetic data generation~\citep{Nikolenko:2019:SDD}. One way to accomplish synthetic data generation is engineering with domain knowledge a reliable model from which to sample data points, e.g., creating computer 3D models to sample images~\citep{Aubry:2015:UDF}. Learning generative models from a small initial dataset is another way to establish a source of synthetic data. Two prominent approaches towards machine learning generative models are generative adverserial networks~\citep{Goodfellow:2014:GAN} and variational autoencoders~\citep{Diederik:2014:AEV}.

The distinction between synthetic and augmented data can become ambiguous in some cases, as augmented data means taking data from real measurements and changing the data in some way that preserves key qualities of the data point while challenging the model to learn the preserved relationships. On the one hand ``synthetic'' data could be considered to include all data that are not the result of direct measurement\footnote{McGraw-Hill Dictionary of Scientific and Technical Terms. Retrieved November 29, 2009.}
, which would subsume data augmentation. On the other, ``synthetic data'' implies that the data is constructed to represent an underlying distribution beyond simply transforming original data with certain invariances. Following the above distinction, DBNQA~\citep{Hartmann:2018:GLD} may represent an ambiguous case, as semantically, the generated instances are all novel as in synthetic data, but syntactically, they are variations that serve to reinforce the shared pattern, as in data augmentation. Our work hopefully elucidates this further.  

Generating synthetic data to train machine learning models has been done for a large number of tasks. Within the field of information retrieval (IR), synthetic data generation has been explored to train models for various tasks, including ad hoc document retrieval~\citep{Azzopardi:2007:BSQ}, suggesting NLQs to clarify search intent from query terms~\citep{Ding:2018:GSD}, and query auto-completion~\citep{Krishnan:2020:GSQ}.

Synthetic data generation has also been used for question answering (QA) tasks~\citep{Zhang:2019:ASD, Golub:2017:TSS, Alberti:2019:SQA, Yang:2019:EEG}. 
Much effort has focused on the machine reading comprehension (MRC) variant of QA, where questions should be answered in the context of a prose paragraph. For example, \citet{Golub:2017:TSS} looked at how to improve transfer learning, fine-tuning a model (pre-trained on one source domain MRC dataset) with synthetic MRC data generated from the target domain corpus of context paragraphs. The common approach, also taken by \citet{Alberti:2019:SQA}, is to use neural language models to select answer spans from paragraphs, and to generate questions conditioned on the answer and paragraph.

We focus on one particular flavor of QA, KGQA, and the datasets and synthetic data generation approaches for this task are discussed in more detail in Sect.~\ref{sec:kgqa-datasets}. 

%% file: ictir2020-synthetic-03.tex
\section{KGQA Datasets and Approaches}
\label{sec:kgqa}

\begin{table*}[t]
\caption{Datasets for complex KGQA.}
\label{tab:datasets}
\vspace*{-0.5\baselineskip}
\begin{tabular}{lll p{3cm} p{5cm}}
 \toprule
 \textbf{Dataset} & \textbf{KG} & \textbf{Size} & \textbf{Generation} & \textbf{Manual post-processing}  \\
 \midrule
QALD-\{1-9\}\footnote{\url{https://project-hobbit.eu/}, \url{https://github.com/ag-sc/QALD}} & DBpedia  & $\sim$50-500 each & Manual & N/A \\ 
QALD-7-train
~\citep{Usbeck:2017:7OC} 
& DBpedia  & 517 & Manual & Programmatically filtered \\ 
LC-QuAD~\citep{Trivedi:2017:LCQ} & DBpedia & 5 000 & Hand-made templates & Paraphrasing$^\dag$ and reviews$^\ddag$ of NLQs \\ 
LC-QuAD 2.0~\citep{Dubey:2019:LCQ} & DBpedia, Wikidata  & 30 000  & Hand-made templates & Paraphrasing$^\dag$ and reviews$^\dag$ of NLQs \\ 
ComplexWebQuestions~\citep{Talmor:2018:WKB}\footnote{\url{http://nlp.cs.tau.ac.il/compwebq}, \url{https://github.com/alontalmor/WebAsKB}}  & Freebase & 34 689 & Hand-made templates & Paraphrasing$^\dag$ of NLQs \\ 
DBNQA~\citep{Hartmann:2018:GLD}\footnote{\url{https://github.com/AKSW/DBNQA}}  & DBpedia & 894 499 & Extracted templates & Reviews$^\ddag$ of generated templates \\
\bottomrule
$^\dag$ Non-experts, $^\ddag$ experts.\\
\end{tabular}
\end{table*}

Knowledge-graph question answering (KGQA) is the task of, given a natural language query $q$, predicting a formal query $f$ that executes on a knowledge graph (KG) $\mathcal{K}$ to return the correct answer $a$, and where $f$ also correctly represents the meaning of $q$. 

In the field of KGQA, a small number of datasets make up the basis for most of the research, as detailed in Sect.~\ref{sec:kgqa-datasets}. Given some dataset, there is a relatively greater variety of approaches to build or train KGQA models, some of which are highlighted in Sect.~\ref{sec:kgqa-approaches}. In this context, the scope of the present work is explained in Sect.~\ref{sec:kgqa-scope}. 

\input{ictir2020-synthetic-03-01.tex}
\input{ictir2020-synthetic-03-02.tex}
\input{ictir2020-synthetic-03-03.tex}

%% file: ictir2020-synthetic-03-01.tex
\subsection{Datasets}
\label{sec:kgqa-datasets}

From 2013 and onwards, KGQA research was conducted on datasets typically on a scale of hundreds of data points or more~\citep{Cai:2013:LSP, Berant:2013:SPF, Yih:2015:SPS, Bao:2016:CBQ, Su-2016-GCQ, Bordes:2015:LSQ, Serban:2016:GFQ}. However, a number of these datasets were exclusively or primarily aimed at simple KGQA~\citep{Bordes:2015:LSQ, Serban:2016:GFQ}. As part of the chosen scope in the present work, we consider only datasets for complex KGQA.\footnote{While some of the excluded KGQA datasets, such as WebQuestions~\citep{Yih:2015:SPS} and WebQuestionsSP~\cite{Yih:2016:VSP}, may include some complex queries, the majority of their queries were simple.} We further consider only those KGQA datasets where each data point consists of a question-query pair, in other words a natural language question (NLQ) and a logical form or formal query representing the NLQ with respect to the KG. The relevant datasets for complex KGQA found in recent literature (since 2013) are summarized in Table~\ref{tab:datasets}. In some cases, the complex KGQA datasets were constructed from simple KGQA datasets, such as ComplexWebQuestions~\citep{Talmor:2018:WKB}, which was constructed from WebQuestionsSP~\citep{Yih:2016:VSP}. In the following, these datasets are described in a bit more detail. 

A series of datasets, from the Question Answering over Linked Data (QALD) challenges\footnote{\url{https://project-hobbit.eu/}, \url{https://github.com/ag-sc/QALD}}, were almost exclusively created manually at small scale. Within this initiative, the re-use and revision of data from previous years has been common. Out of the various QALD datasets, QALD-7-train~\citep{Usbeck:2017:7OC} is highlighted due to its use among the seed data from which \citet{Hartmann:2018:GLD} extracted templates.

The LC-QuAD~\citep{Trivedi:2017:LCQ} dataset was created from a set of 38 \footnote{The published file only contains 35 templates.} hand-made abstract query subgraphs extending at most two hops from a seed entity. These were instantiated with whitelisted entities and predicates, and a template for expressing the query as an NLQ was correspondingly populated. This tentative template-based NLQ was paraphrased by crowdsourced non-experts to improve the grammar of the question. The resulting paraphrased NLQs were then reviewed and revised by experts.  

The LC~QuAD 2.0~\citep{Dubey:2019:LCQ} dataset was created in a similar manner, except the initial set of 22 
query subgraph templates were constructed not from geometric constraints but from consideration of pre-existing QA datasets. The template-based NLQs were paraphrased by crowdsourcing non-experts, the results were likewise paraphrased, and a third round of crowdsourcing verified whether or not the two paraphrases of the NLQ were identical in meaning. 

The ComplexWebQuestions~\citep{Talmor:2018:WKB} dataset was created by taking the simple KGQA question-query pairs from WebQuestionsSP~\citep{Yih:2015:SPS} and constructing templates to add constraints to each data point to produce complex KGQA data. The NLQs were extended with manually constructed predicate-specific templates. Tentative template-based NLQs were then paraphrased by crowdsourced non-experts. 

Finally, the DBNQA~\citep{Hartmann:2018:GLD} dataset was constructed by extracting templates of paired NLQ and SPARQL queries, one from each seed data point selected from QALD-7-train~\citep{Usbeck:2017:7OC} and LC-QuAD \cite{Trivedi:2017:LCQ}. The templates derived from QALD-7-train were extracted manually. Meanwhile, templates could be extracted semi-automatically from LC-QuAD: first a script exploited the indicated surface forms in the NLQs, and then the resulting templates were reviewed by SPARQL experts. For each entity URI or surface form in the seed data, corresponding placeholders were inserted in the templates. The templates were then instantiated using the results of the executable SPARQL templates applied to a DBpedia endpoint to find entities for the placeholders.  

The datasets listed in Table~\ref{tab:datasets} indicate a trend towards scalable instance generation to economically generate larger volumes of question-query pair data for complex KGQA. The progression started with fully manual dataset generation, moving to using hand-made templates for automated instance generation, and now with DBNQA the automated template generation from pre-existing seed datasets. While these changes in automation have increased the scale of available datasets, the need for manual post-processing also increases. Consequently, it becomes economically desirable to divide the post-processing work into (i) work that requires expert knowledge of the formal query language, and (ii) work that can be adequately performed by crowdsourced non-experts. The non-experts only need adequate natural language skills to ameliorate the grammatical artifacts of template-based NLQ generation. 

%% file: ictir2020-synthetic-03-02.tex
\subsection{Approaches}
\label{sec:kgqa-approaches}

Various approaches have been taken for developing complex KGQA systems. One important distinction is between neural and non-neural approaches. The former represents a recent trend, focusing on the development of suitable neural architectures for end-to-end learning, while the latter tends to decompose the KGQA task into a sequence of discrete subtasks and create purpose-built solutions for each.

\subsubsection{Non-neural Approaches}

As if to indicate the pervasive shift towards neural approaches, \citet{Chakraborty:2019:INN} refer to non-neural KGQA approaches as ``traditional''~\citep{Berant:2013:SPF, Unger:2014:QAL, Reddy:2014:LSS}. \citet{Diefenbach:2017:CTQ} consider all KGQA tasks to consist of distinct stages, all of which must be solved by the KGQA system.  
For example, the message-passing architecture QAmp~\cite{Vakulenko:2019:MPC} can be considered a hybrid of neural and non-neural approaches, but is not an end-to-end neural system. QAmp uses neural components (RNN classifiers, word embeddings) in its question interpretation stage, in addition to index-based methods. In the answer inference stage, the components are using probabilistic graphical models approaches. 
Another hybrid approach was devised by~\citet{Abujabal:2017:ATG}, who separated the semantic parsing for KGQA into various components: an offline neural component that learned from KGQA data to generate the syntactic template components that were used compositionally to parse the semantics of an NLQ into a formal query; another component that generates candidate queries; and a component to rank the candidate queries to output the inferred formal query.

\subsubsection{Neural Architectures}

Perhaps the most common way neural networks are used end-to-end in KGQA is to treat the formal query lanuage, e.g., SPARQL, as a target language in neural machine translation (NMT). As KGQA is cast as a semantic parsing task, this makes sense, although the strict syntax of the formal query language differs from the more varied and flexible syntax in natural languages. 

As shown by~\citet{Yin:2019:NMT}, a number of NMT KGQA architectures can be successfully trained on KGQA data, i.e., NLQ and SPARQL query pairs. Besides the architectures mentioned in Sect.~\ref{sec:kgqa-scope}, \citet{Yin:2019:NMT} tested the following architectures on KGQA datasets, including DBNQA:
\begin{itemize}
	\item Two variants of Google's NMT architectures from~\citet{Wu:2016:GNM} were tested, \emph{GNMT-4} and \emph{GNMT-8}, which are respectively 4- and 8-layer LSTM-based RNNs with a bi-directional encoding layer.
	\item \emph{LSTM\_Luong}, the LSTM-based 4-layer RNN with local attention,  introduced by~\citet{Luong:2015:EAA} was also tested. 
	\item A single CNN-based architecture was tested, \emph{ConvS2S}~\citep{Gehring:2017:CSS}. 
	\item The \emph{Transformer}~\citep{Vaswani:2017:AIA} architecture was also tested. 
\end{itemize}

Neural KGQA is surveyed in greater detail by~\citet{Chakraborty:2019:INN}, including classification and ranking approaches, especially for simple KGQA, as well as machine translation approaches, which may be more suitable for complex KGQA. Various neural machine translation (NMT) approaches to KGQA have been investigated in previous work~\citet{Yih:2016:VSP, Liang:2013:LDC, Yih:2015:SPS, DongLapata:2016:LLF, JiaLiang:2016:DRN, Soru:2017:SFL, Soru:2018:NMT}.

%% file: ictir2020-synthetic-03-03.tex
\subsection{Scope}
\label{sec:kgqa-scope}

Out of the available approaches to KGQA, we consider only neural machine translation (NMT) architectures, which are interpreted as performing \emph{semantic parsing} on the NLQ $q$ to produce a semantically equivalent formal (SPARQL) query $f$ that also executes on the target knowledge graph $\mathcal{K}$, retrieving the correct answer $a$. 

In the present work, we limit ourself to a particular baseline architecture and its variants, to ensure the comparability of the obtained results.  We note that the same experiments can be performed with additional architectures in the future.
Specifically, the baseline architecture is taken from~\citep{Yin:2019:NMT} (originally from \citep{Soru:2017:SFL, Soru:2018:NMT}), as well as two attention-based variations of this architecture, which in \citep{Yin:2019:NMT} performed well on DBNQA. 
The work of \citet{Yin:2019:NMT} was taken as a starting point because it was the only work that considered a variety of NMT architectures applied to the largest available complex KGQA dataset, DBNQA~\citep{Hartmann:2018:GLD}. 
The selection was made both due to the high performance on the randomly partitioned DBNQA, as well as the fact that these models were implemented in the same framework, Tensorflow. 

\begin{description}
	\item[NSpM baseline] The baseline architecture is here referred to as \textbf{NSpM baseline} following \cite{Yin:2019:NMT, Soru:2017:SFL, Soru:2018:NMT}. However, it is a basic Tensorflow NMT architecture, with 2 layers, 128 units per layer, a dropout rate of $20\%$, and optimizing on the BLEU metric.
	\item[NSpM+Att1] The second architecture is called \textbf{NSpM+Att1}, again following \cite{Yin:2019:NMT}, and it differs from NSpM baseline only in that a global Bahdanau attention mechanism is added \cite{Bahdanau:2015:NMT}. Since the type of global Bahdanau attention mechanism utilized in \cite{Yin:2019:NMT} was not further specified, the present work selected a ``normed'' variant.
	\item[NSpM+Att2]The third architecture is called \textbf{NSpM+Att2}, again following \cite{Yin:2019:NMT}, and it differs from NSpM baseline only in that a local Luong attention mechanism is added \cite{Luong:2015:EAA}. Since the type of local Luong attention mechanism utilized in \cite{Yin:2019:NMT} was not further specified, the present work selected a ``scaled'' variant. This architecture had the second-best performance of the 8 architectures evaluated in \cite{Yin:2019:NMT}, second only to the Convolutional sequence-to-sequence architecture \textbf{ConvS2S}, and even that difference was relatively slight.
\end{description}

%% file: ictir2020-synthetic-04.tex
\section{Methodology}
\label{sec:methodology}

\begin{table*}[t]
\caption{Examples of seed, template, and instance.}
\begin{tabular}{lll}
\toprule
	\textbf{Seed} & 
	NLQ & \emph{Is Peter Piper Pizza in the pizza industry?} \\
	& SPARQL & \texttt{ASK WHERE \{<http://dbpedia.org/resource/Peter\_Piper\_Pizza>} \\
	& & \texttt{<http://dbpedia.org/ontology/industry> <http://dbpedia.org/resource/Pizza>\}} \\
\midrule
	\textbf{Template} & 
	NLQ & \emph{Is <B> in the <A> industry?} \\
	& SPARQL & \texttt{SELECT DISTINCT ?a, ?b WHERE \{?b <http://dbpedia.org/ontology/industry> ?a\} }\\
\midrule
	\textbf{Instance 1} & 
	NLQ & \emph{Is robot comics in the publishing industry?} \\
	& SPARQL & \texttt{ASK WHERE \{<http://dbpedia.org/resource/Robot\_Comics>} \\
	& & \texttt{<http://dbpedia.org/ontology/industry> <http://dbpedia.org/resource/Publishing>\}} \\
\midrule
	\textbf{Instance 2} & 
	NLQ & \emph{Is tiger aircraft in the aerospace industry?} \\
	& SPARQL & \texttt{ASK WHERE \{<http://dbpedia.org/resource/Tiger\_Aircraft>} \\
	& & \texttt{<http://dbpedia.org/ontology/industry> <http://dbpedia.org/resource/Aerospace>\}} \\
\bottomrule
\end{tabular}
\label{tab:examples}
\end{table*}

We are looking at a dataset~\citep{Hartmann:2018:GLD} where instances were generated with templates extracted from seeds~\citep{Trivedi:2017:LCQ, Usbeck:2017:7OC}. The evaluation of this synthetic dataset was done with randomly partitioned training, validation, and test splits~\citep{Yin:2019:NMT}. This random partitioning did not avoid allocating instances generated from the same template to different splits. Thus, models trained and evaluated on this random partitioning would see ``familiar'' instances in the validation and test splits, i.e., instances generated from the same template as instances used for training that model. Could this have created an information leakage, whereby the trained models have memorized a finite set of underlying templates---those \emph{seen} during training---rather than learning to generalize from training instances to previously \emph{unseen} patterns? 

To answer this question, we have designed a method to \emph{sanitize}\footnote{This term might be value-laden and undescriptive of the mechanics employed, but reflects the intention to remove or minimize and contamination due to information leakage.} the existing dataset, and ensure that a held-out test split contains only instances generated from a held-out split of templates. 
Since we are working with a pre-existing dataset with no labelling or index of which template generated each instance, the sanitation process is inevitably somewhat uncertain and depends on constructing a reasonable set of rules to identify which template generated an instance.  We make a best effort to recover this information, that is, the originating template for each instance. 

\begin{figure}[t]
    \centering
	\includegraphics[width=0.4\textwidth]{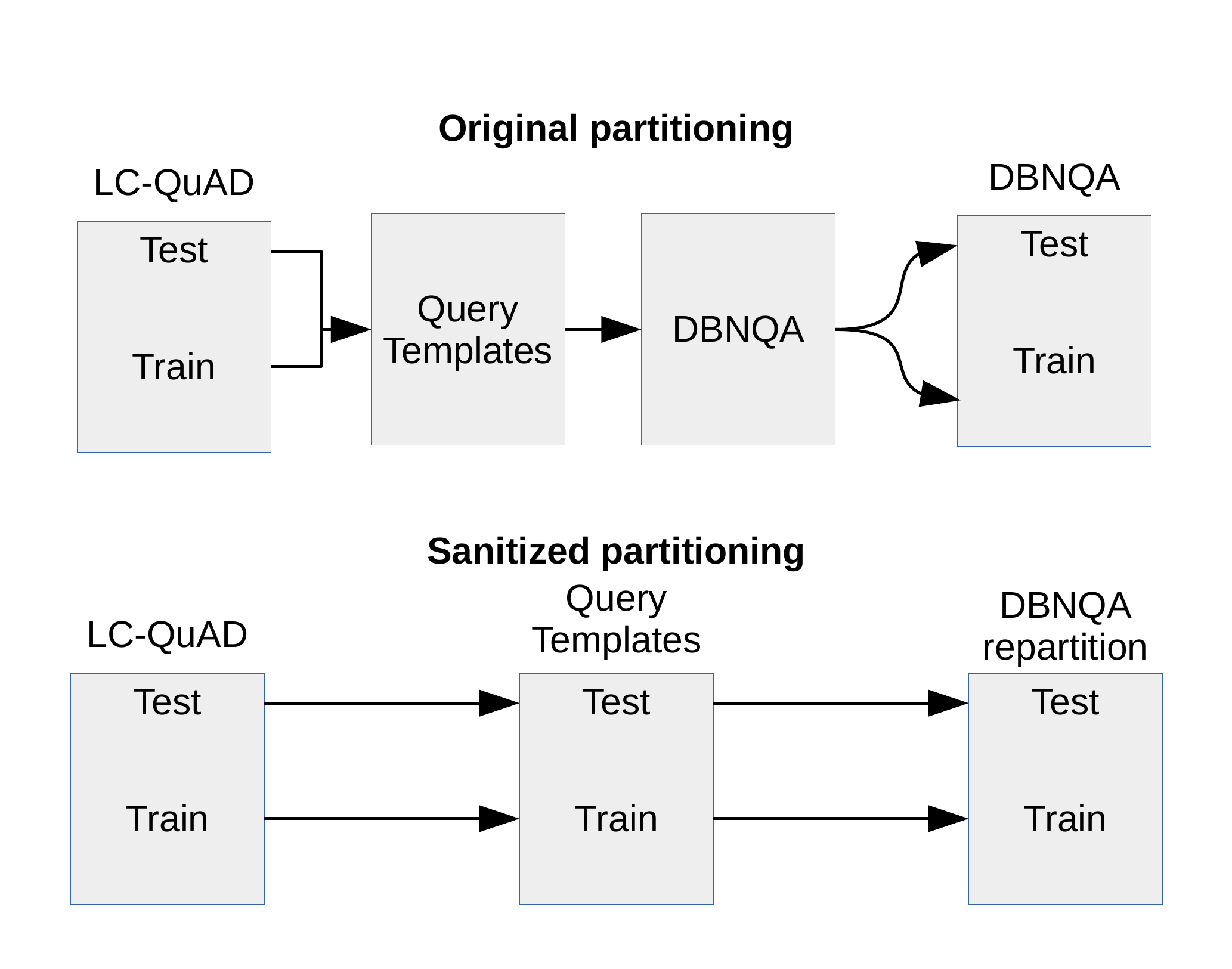}
	\caption{Illustration of original and sanitized DBNQA partitioning.}
	\label{fig:partitions}
\end{figure}

We can consider that the two approaches taken, shown in Fig.~\ref{fig:partitions}, the fully random partitioning by~\citet{Yin:2019:NMT}, and the sanitized partitioning in the present work, may represent two extremes in how template-based synthetic data should be treated. This perspective is further developed in Sect.~\ref{sec:concl}. 

\subsection{Preliminaries}

The pipeline utilized by~\citet{Hartmann:2018:GLD} to generate a large volume of KGQA training data can be considered as three discrete stages, which we refer to as \emph{seeds}, \emph{templates}, and \emph{instances}. First, a small high-quality KGQA dataset consisting of question-query pairs is taken as the seed dataset $s \in \mathcal{S}$ from which are extracted templates $t \in \mathcal{T}$, capturing the underlying pattern of the seed data points, but replacing certain parts of the seed data points with placeholder tokens or URIs, for NLQ and SPARQL forms, respectively. The templates are then instantiated into concrete data points, replacing the placeholders in a template with appropriate terms (entity labels) or entity URIs. Each template can be used to generate an arbitrary number of such new instances $i \in \mathcal{I}$, bounded only by the availability of unique paths (subgraphs) on the knowledge graph that fit the path(s) of the template. The different stages are illustrated in Table~\ref{tab:examples}, with a pair of NLQ and SPARQL forms for each stage. The examples are chosen such that the template is derived from the seed, and the instances are both generated from the same template. Two example instances are shown to illustrate the similarities of instances generated from the same template.

\subsection{Original DBNQA}
\label{sec:methodology-original}
\begin{table}[t]
\caption{Overview of dataset splits used in our experiments. Five different random splits of original DBNQA were used with these proportions. Sanitized-1 DBNQA and Sanitized-2 DBNQA were based on 20\% and 10\% test splits in the LC-QuAD seed set, respectively.}
\begin{tabular}{@{}lr@{~}l@{~~~}r@{~}l@{~~~}r@{~}l@{}}
	\toprule	
 	\textbf{Dataset} & \multicolumn{2}{c}{\textbf{Train}} & \multicolumn{2}{c}{\textbf{Validation}} & \multicolumn{2}{c}{\textbf{Test}} \\
	\midrule
	Original DBNQA & 715 600 & (80.0\%) & 89 449 & (10.0\%) & 89 450 & (10.0\%) \\
	Sanitized-1 DBNQA & 659 313 & (74.8\%) & 73 257 & (8.3\%) & 148 397 & (16.8\%) \\
	Sanitized-2 DBNQA & 726 355 & (82.4\%) & 80 706 & (9.2\%) & 73 906 & (8.4\%) \\
	\bottomrule
\end{tabular}
\label{tab:repart}
\end{table}

The DBNQA dataset is provided without any canonical partitions~\citep{Hartmann:2018:GLD}. Researchers are free to randomly partition the dataset into training, validation, and testing splits.  This was done by~\citet{Yin:2019:NMT}, who reported allocating $80\%$-$10\%-10\%$ to the respective splits. However, their unique partitioning is not recoverable from their paper or code repository. The present work randomly partitioned DBNQA in the same proportions, but used specific random seeds. In order to ensure that any differences were not due to random chance, this random partitioning was done five times with different random seeds each time, and the resulting partitions were used in Experiment 1 (see Sect.\ref{sec:experiments:exp1}) to separately train models of each of the three  architectures discussed in Sect.~\ref{sec:kgqa-approaches}.

\subsection{Sanitized DBNQA}
\label{sec:methodology-repartitioned}

In order to investigate the question of information leakage via templates, the train-and-test-splits partitioning of the major part of the seed dataset, LC-QuAD, was used to coordinate a partitioning of the LC-QuAD-based templates dataset previously used in generating DBNQA. Subsequently, the partitioned templates were used to partition the instances dataset, DBNQA. This repartitioning is illustrated in the bottom half of Fig.\ref{fig:partitions}.
Templates were assigned to the template test split $\mathcal{T}_\text{test} \subset \mathcal{T}$ if the NLQ forms of both the seed and template were identical except where the template placeholders allow a contiguous sequence of tokens in the seed, and if the predicates in the seed SPARQL are the same as those in the template SPARQL. Similarly, instances were assigned to the instance test split $\mathcal{I}_\text{test}$ if the NLQ forms match as above, and if all the predicates in the template SPARQL are also in the instance SPARQL, in the same order.

The datasets were also de-duplicated at each stage. The original and sanitized instance datasets are summarized in Table~\ref{tab:repart}. Sanitized-1 DBNQA was based on the canonical split of LC-QuAD into an 80\% training split and a 20\% test split. Sanitized-2 DBNQA was based on a 90\%-10\% split of LC-QuAD. 
Only after repartitioning in this systematic manner based on template matching is the instance training split itself randomly partitioned into a $90\%$ training split and a $10\%$ validation split. Thus, the test split is sanitized with respect to the training split, while the validation split is not. By evaluating trained models on both the test split and validation split, we illustrate the information leakage via templates caused by a purely random partitioning of an instance dataset like DBNQA. 
Having thus repartitioned the instances, we trained KGQA NMT models as in~\citet{Yin:2019:NMT}, both reproducing the approach taken by Yin, et al. with the original DBNQA partitionings described in Sect.~\ref{sec:methodology-original}, as well as on the repartitioned datasets, to compare the results of training with and without information leakage across the dataset splits. 

%% file: ictir2020-synthetic-05.tex
\section{Experiments}
\label{sec:experiments}

\begin{table*}[t]
\caption{Results of comparing original DBNQA and Sanitized-1 DBNQA to train and evaluate models.}
\begin{tabular}{lcccccccc}
\toprule
  & \multicolumn{4}{c}{\textbf{Original DBNQA}} & \multicolumn{4}{c}{\textbf{Sanitized-1 DBNQA}} \\
  \textbf{Architecture} & \multicolumn{2}{c}{\textbf{BLEU}} & \multicolumn{2}{c}{\textbf{Perplexity}} & \multicolumn{2}{c}{\textbf{BLEU}} & \multicolumn{2}{c}{\textbf{Perplexity}} \\
  & Valid. & Test & Valid. & Test & Valid. & Test & Valid. & Test \\
 \midrule
NSpM baseline & $62.56 \pm 0.10$ & $62.52 \pm 0.10$ & $2.37 \pm 0.01$ & $2.37 \pm 0.01$ & 64.96 & 41.12 & 2.33 & 11.86 \\
NSpM+Att1     & $79.27 \pm 1.82$ & $79.22 \pm 1.77$ & $1.58 \pm 0.06$ & $1.58 \pm 0.06$ & 85.08 & 54.39 & 1.60 & 7.73 \\
NSpM+Att2     & $80.58 \pm 0.95$ & $80.53 \pm 0.89$ & $1.54 \pm 0.03$ & $1.54 \pm 0.02$ & 84.67 & 54.55 & 1.61 & 9.01 \\
\bottomrule
\end{tabular}
\label{tab:results-exp1}
\end{table*}

\begin{figure*}[h!]
   \centering
   \begin{tabular}{c@{}c@{}c@{}c@{}}
   \includegraphics[width=.25\textwidth]{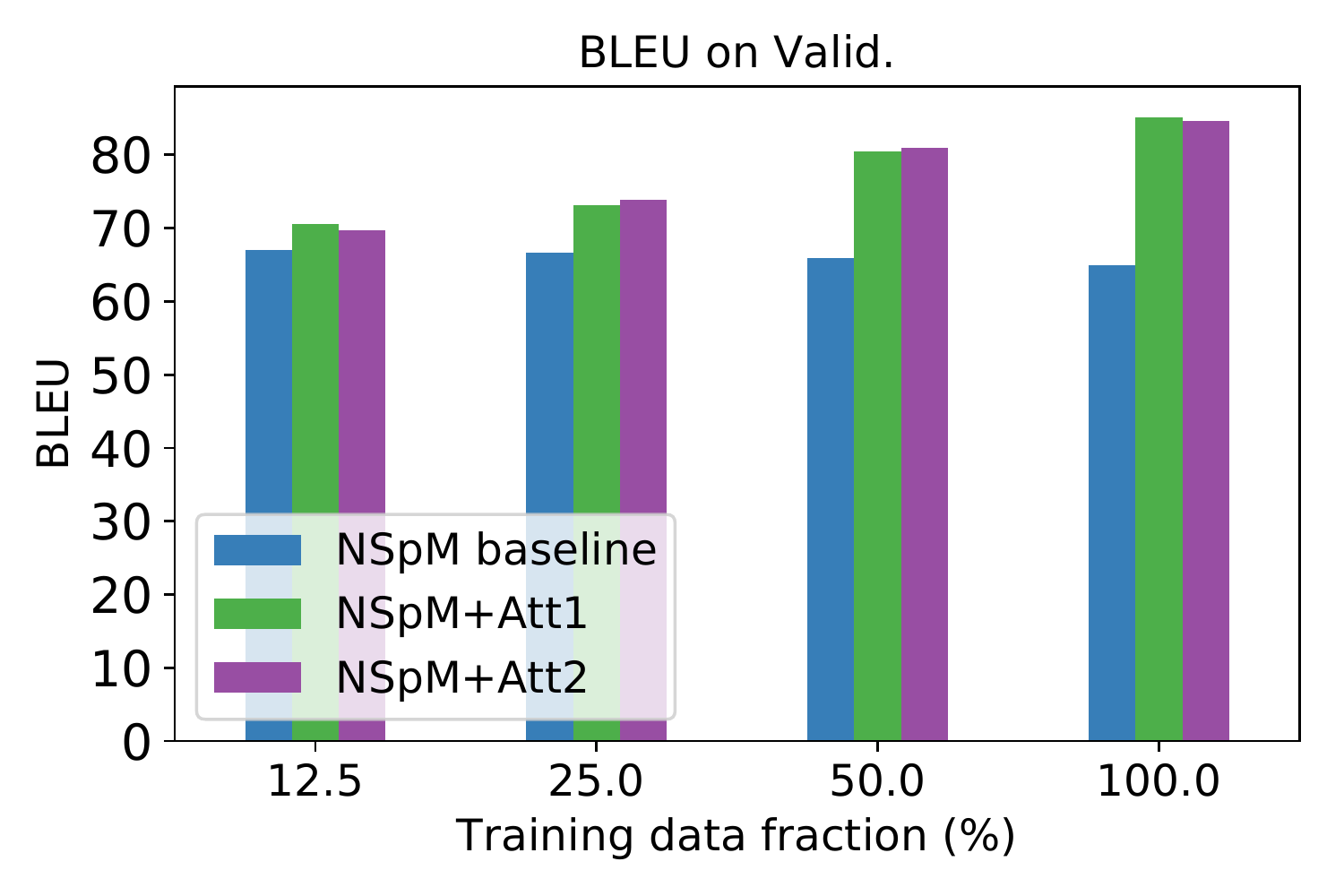} & 
     \includegraphics[width=.25\textwidth]{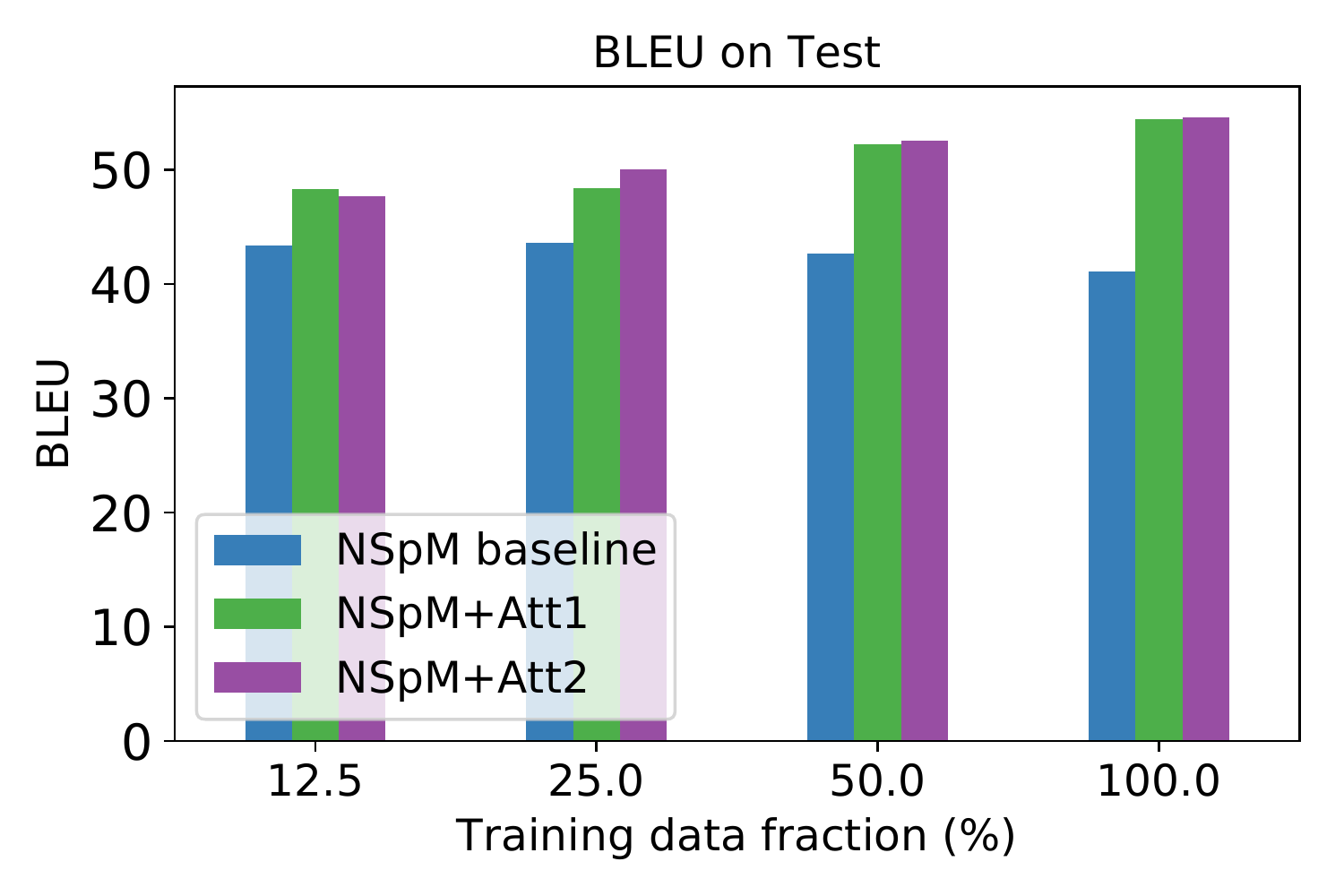} & 
    \includegraphics[width=.25\textwidth]{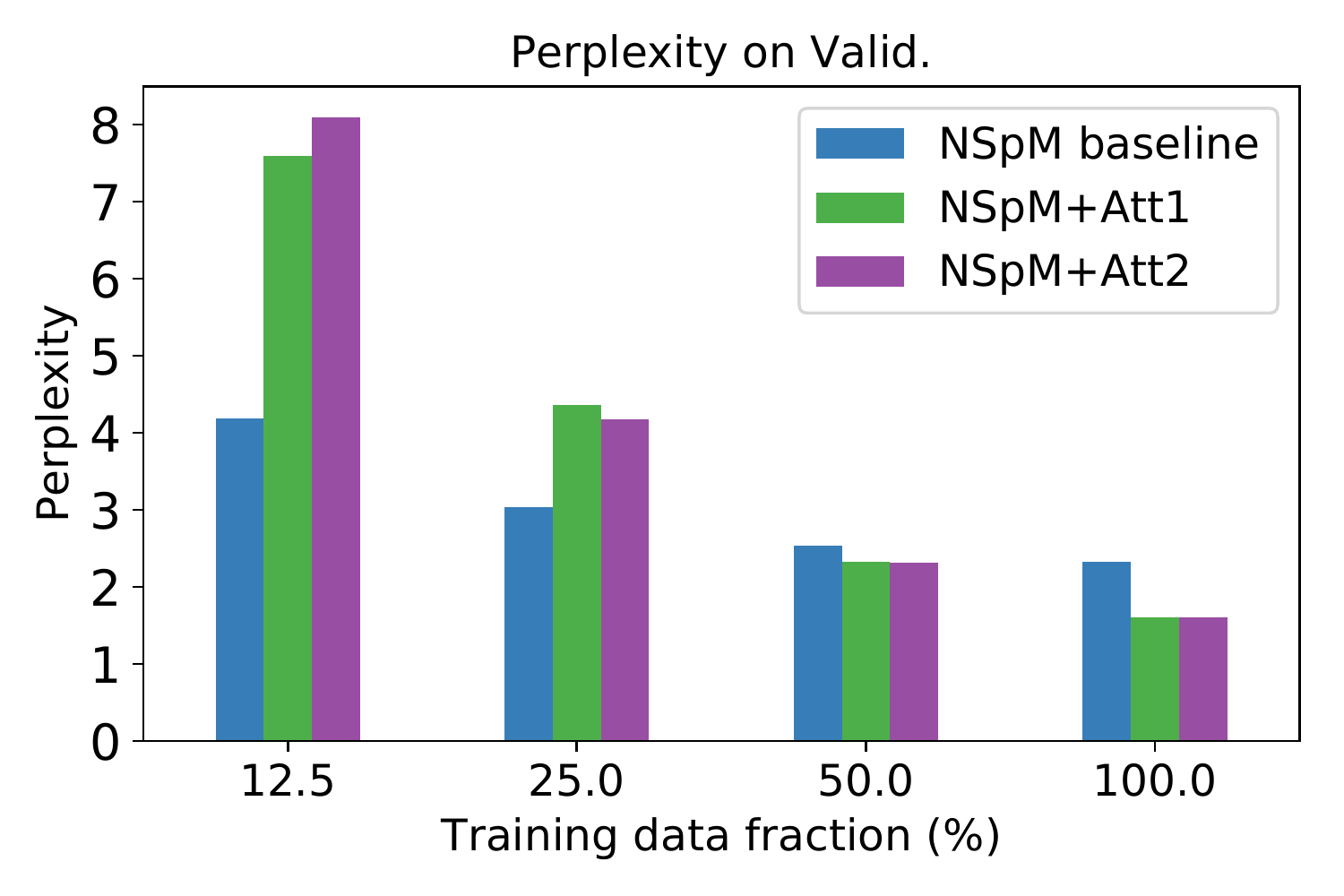} & 
    \includegraphics[width=.25\textwidth]{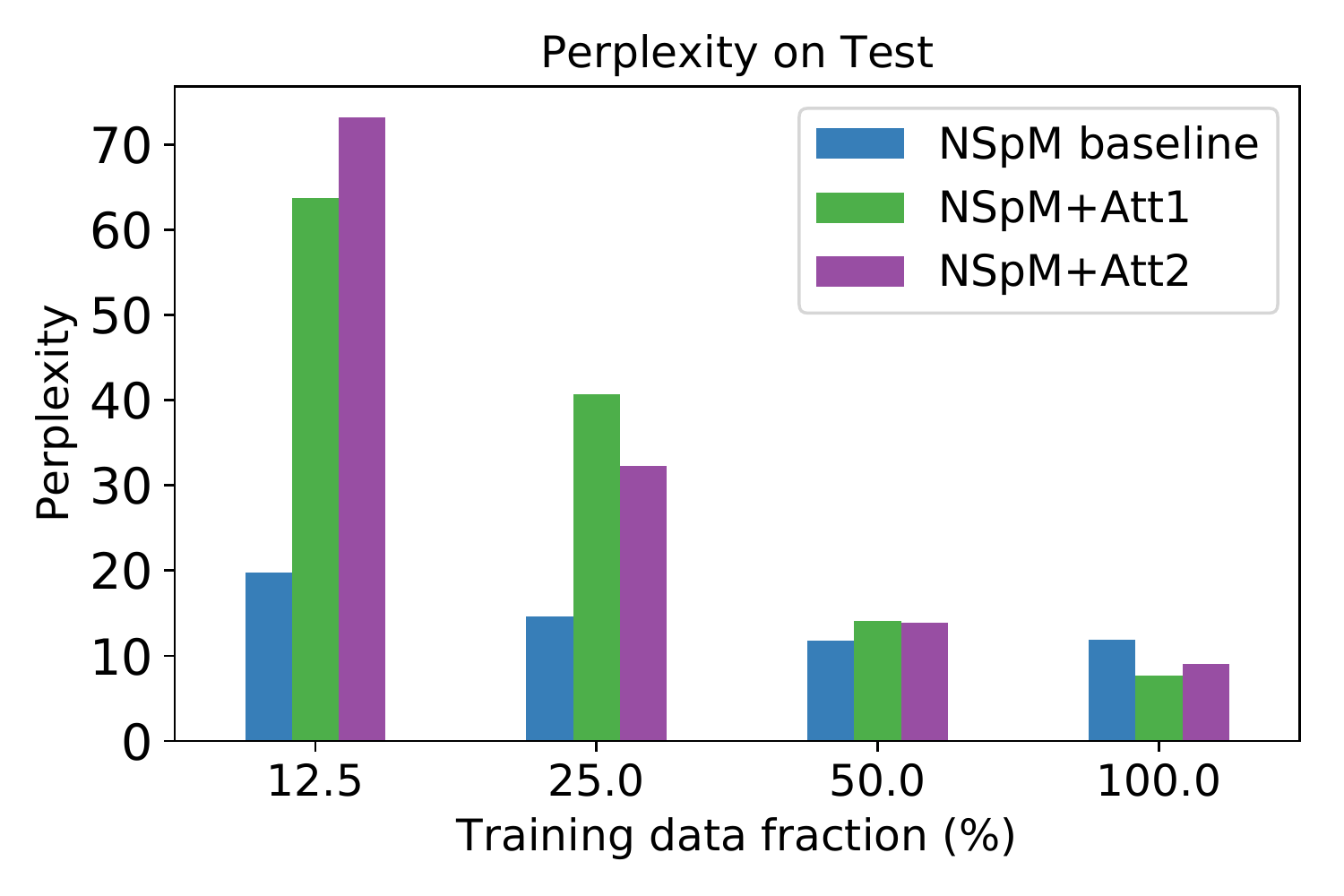}
   \\      
   \end{tabular}    
\caption{Results of comparing models trained on different fractions of the Sanitized-1 DBNQA training split.}
   \label{fig:results-exp2}
\end{figure*}

This section presents a series of experiments we performed on the original and sanitized DBNQA collections, to answer our research questions. Model performance is evaluated in terms of the metrics BLEU~\citep{Papineni:2002:BLEU} and perplexity~\citep{Brown:1992:EUB}, for comparison with results reported by~\citet{Yin:2019:NMT}. These measures are commonly used in machine translation evaluation, especially BLEU, which reflects how well the predicted output matches with the ground truth. Perplexity reflects the degree of surprise caused by the model's predictions compared to the ground truth. 

\subsection{Sanitized data partitioning}
\label{sec:experiments:exp1}
Is the performance of trained models affected by the sanitized data partitioning? 
We addressed this question with our first experiment, which compared models trained on either the original DBNQA, or on Sanitized-1 DBNQA, which was partitioned based on the canonical partitioning of LC-QuAD as described in Sect.~\ref{sec:methodology-repartitioned}. 
The original DBNQA was randomly partitioned five times with unique random seeds but identical proportions between the splits ($80\%-10\%-10\%$), and the models were independently trained on each random partition. The performance results of original DBNQA are therefore shown as the mean and $\pm$ standard deviation of the models across the five random partitionings. 
These results, listed in Table~\ref{tab:results-exp1}, show  that the performance of the trained models was very similar across the test and validation splits of original DBNQA, as well as the unsanitized validation split of Sanitized-1 DBNQA. However, the sanitized test split showed a significant reduction in performance across all the trained models, in terms of both BLEU and perplexity. 

\subsection{Varying volumes of training data}
\label{sec:experiments:exp2}

Is the performance of the trained models with respect to the sanitized test split dependent on the amount of training data? 
This would indicate whether there is some degree of generalization from instances based on templates that have been \emph{seen} before during training, or if there is no discernible generalization at all. We addressed this question with our second experiment, which compared the effects of different amounts of training data on the trained models' performance with respect to the Sanitized-1 DBNQA validation and test splits. Although performance increase as a function of increased training data volume is the expected behavior, it has been shown to not always be the case in QA~\citep{Linjordet:2019:ITD}.  Thus, it is not \emph{a priori} certain that the sanitized test split is similar enough to the sanitized training split that the expected behavior occurs. This experiment verifies whether or not what the models learn generalizes to the sanitized test split \emph{proportionally} to the volume of training data. Thus, the experiment also elucidates whether the models are only learning to memorize the \emph{seen} template patterns, or is learning to generalize to instances from \emph{unseen} templates.

We trained with fractions of the training data used in our first experiment: $12.5\%$, $25\%$, $50\%$, and $100\%$.
From the results shown in Fig.~\ref{fig:results-exp2}, there is a clear trend for the models NSpM+Att1 and NSpM+Att2, where increased amounts of training data yield improved model performance on both the sanitized test split and the unsanitized validation split, in terms of both BLEU and perplexity. The NSpM baseline model, on the other hand, does not benefit as much from increased training data, improving in terms of perplexity, but even deteriorating slightly in terms of BLEU. However, this also holds for the unsanitized validation split, and so reflects on the architecture's ability to improve from training data, not on the sanitized partitioning. 

For all models, performance is reduced by the challenge of the sanitized test split, but where performance improves with increased training data, it does so even on the sanitized test split. 

\subsection{Varying size of seed partitions before sanitation}
\label{sec:experiments:exp3}

\begin{table}[t]
\caption{Results of evaluating models trained on Sanitized-2 DBNQA.}
\begin{tabular}{lcccc}
\toprule
  \textbf{Architecture} & \multicolumn{2}{c}{\textbf{BLEU}} & \multicolumn{2}{c}{\textbf{Perplexity}} \\
  & Valid. & Test & Valid. & Test \\
 \midrule
NSpM baseline & 64.35 & 42.58 & 2.36 & 10.01 \\ 

NSpM+Att1     & 82.87 & 53.09 & 1.59 & 7.33  \\ 
NSpM+Att2     & 86.05 & 56.94 & 1.52 & 6.57 \\ 
\bottomrule
\end{tabular}
\label{tab:results-exp3}
\end{table}

Is it possible that different partitions of the seed set can affect the degree of generalization? 
To address this question, in our third experiment we investigated whether increasing the proportion of templates \emph{seen} via the training instances would translate into improved performance on the sanitized test split. We divided the canonical test split of LC-QuAD in half, and added one half back into the seed training split, before doing the sanitizing procedure, yielding Sanitized-2 DBNQA, as described in Sect.~\ref{sec:methodology-repartitioned}. 

As can been seen from results shown in Table~\ref{tab:results-exp3}, here as in our second experiment the performance of all models on the unsanitized validation split was generally better than the performance of models trained on the original DBNQA. Models trained on Sanitized-2 DBNQA performed similarly to models trained on Sanitized-1 DBNQA, with some variations on the order of the standard deviations seen for original DBNQA in Table~\ref{tab:results-exp1}. We note, however, that all models trained on Sanitized-2 DBNQA performed better in terms of perplexity with respect to the sanitized test split. 

%% file: ictir2020-synthetic-06.tex
\section{Discussion and Conclusion}
\label{sec:concl}

As we have seen from the datasets presented in Sect.~\ref{sec:kgqa-datasets}, there is a trend towards satisfying an important desideratum of machine learning generally, and deep learning in particular: (i) a large-scale training dataset of such quality and variety that it allows the model to observe and learn to predict patterns when presented new data from the same distribution.
However, there is an important second desideratum: (ii) a test set that comprises data from the same distribution but which is novel enough to the model so that model performance is due to the model learning the underlying dynamics of the data, rather than memorizing a finite set of patterns. 

In the present work, we have questioned whether DBNQA as used in previous work~\citep{Yin:2019:NMT} satisfies (ii). 
Our hypothesis is that there is a leakage of information between the DBNQA training split on the one hand and the validation and test splits on the other, as used by~\citet{Yin:2019:NMT}.  
We argue that~\citet{Yin:2019:NMT} have sacrificed (ii) in favor of (i), while in this paper we considered the other extreme, where (i) is sacrificed in favor of (ii). For future work, we speculate, is there a middle ground that can be reliably found?

In our experiments, we first showed in Sect.~\ref{sec:experiments:exp1} that there is indeed a large difference in performance on the test split of our sanitized DBNQA partitioning, compared to the validation split, which is randomly partitioned in a template-naive manner. From our second experiment in Sect.~\ref{sec:experiments:exp2}, we showed that for models that improve with increased volumes of training data, that improvement also generalizes to the sanitized test split. Finally, in our third experiment in Sect.~\ref{sec:experiments:exp3}, the models trained on Sanitized-2 DBNQA showed some tendency to improve performance on both validation and test split, indicating generalization from \emph{seen} to \emph{unseen} templates. 

Our results raise a set of interesting questions around training models with synthetic data using fair conditions. These are questions raised by the present study that may be the subject for future work: 
How well do these findings generalize to other model families than those tested here? Of particular interest are the architectures of ConvS2S~\citep{Gehring:2017:CSS}, Transformer~\citep{Vaswani:2017:AIA}, and BERT~\citep{Devlin:2019:BERT}. Can the distinction between memorization and generalized learning be precisely characterized? How can synthetically generated training data be structured to promote learning dynamics (e.g., of a formal syntax) rather a finite set of fixed patterns? For template-based synthetic data generation, what should be the relationship between training and test splits to fairly evaluate the performance of trained models? 

In summary, we have shown that several NMT-based neural KGQA systems have reduced performance on instances generated from templates where the models saw no instances generated from those templates during training. At the same time, the performance on instances from such \emph{unseen} templates did show improvement from increased training data, indicating that some models were able to generalize better with more training data. 

We have shown that a significant part of performance in these models as reported by~\citet{Yin:2019:NMT} may largely be attributed to the models learning to recognize the underlying patterns of specific templates from which were generated the instances seen during training.
In contrast, the ideal NMT KGQA system would learn the underlying syntaxes of the source and target languages and handle unseen patterns according to implicit principles.

%% file: 00paper.bbl

\begin{thebibliography}{47}


\ifx \showCODEN    \undefined \def \showCODEN     #1{\unskip}     \fi
\ifx \showDOI      \undefined \def \showDOI       #1{#1}\fi
\ifx \showISBNx    \undefined \def \showISBNx     #1{\unskip}     \fi
\ifx \showISBNxiii \undefined \def \showISBNxiii  #1{\unskip}     \fi
\ifx \showISSN     \undefined \def \showISSN      #1{\unskip}     \fi
\ifx \showLCCN     \undefined \def \showLCCN      #1{\unskip}     \fi
\ifx \shownote     \undefined \def \shownote      #1{#1}          \fi
\ifx \showarticletitle \undefined \def \showarticletitle #1{#1}   \fi
\ifx \showURL      \undefined \def \showURL       {\relax}        \fi
\providecommand\bibfield[2]{#2}
\providecommand\bibinfo[2]{#2}
\providecommand\natexlab[1]{#1}
\providecommand\showeprint[2][]{arXiv:#2}

\bibitem[\protect\citeauthoryear{Abujabal, Yahya, Riedewald, and
  Weikum}{Abujabal et~al\mbox{.}}{2017}]%
        {Abujabal:2017:ATG}
\bibfield{author}{\bibinfo{person}{Abdalghani Abujabal},
  \bibinfo{person}{Mohamed Yahya}, \bibinfo{person}{Mirek Riedewald}, {and}
  \bibinfo{person}{Gerhard Weikum}.} \bibinfo{year}{2017}\natexlab{}.
\newblock \showarticletitle{Automated Template Generation for Question
  Answering over Knowledge Graphs}. In \bibinfo{booktitle}{\emph{Proc. of WWW
  '17}}. \bibinfo{pages}{1191--1200}.
\newblock


\bibitem[\protect\citeauthoryear{Alberti, Andor, Pitler, Devlin, and
  Collins}{Alberti et~al\mbox{.}}{2019}]%
        {Alberti:2019:SQA}
\bibfield{author}{\bibinfo{person}{Chris Alberti}, \bibinfo{person}{Daniel
  Andor}, \bibinfo{person}{Emily Pitler}, \bibinfo{person}{Jacob Devlin}, {and}
  \bibinfo{person}{Michael Collins}.} \bibinfo{year}{2019}\natexlab{}.
\newblock \showarticletitle{Synthetic {QA} Corpora Generation with Roundtrip
  Consistency}. In \bibinfo{booktitle}{\emph{Proc. of ACL '19}}.
  \bibinfo{pages}{6168--6173}.
\newblock


\bibitem[\protect\citeauthoryear{Aubry and Russell}{Aubry and Russell}{2015}]%
        {Aubry:2015:UDF}
\bibfield{author}{\bibinfo{person}{M. Aubry} {and} \bibinfo{person}{B.~C.
  Russell}.} \bibinfo{year}{2015}\natexlab{}.
\newblock \showarticletitle{Understanding Deep Features with Computer-Generated
  Imagery}. In \bibinfo{booktitle}{\emph{Proc. of ICCV '15}}.
  \bibinfo{pages}{2875--2883}.
\newblock


\bibitem[\protect\citeauthoryear{Azzopardi, de~Rijke, and Balog}{Azzopardi
  et~al\mbox{.}}{2007}]%
        {Azzopardi:2007:BSQ}
\bibfield{author}{\bibinfo{person}{Leif Azzopardi}, \bibinfo{person}{Maarten de
  Rijke}, {and} \bibinfo{person}{Krisztian Balog}.}
  \bibinfo{year}{2007}\natexlab{}.
\newblock \showarticletitle{Building Simulated Queries for Known-item Topics:
  An Analysis Using Six European Languages}. In \bibinfo{booktitle}{\emph{Proc.
  of SIGIR '07}}.
\newblock


\bibitem[\protect\citeauthoryear{Bahdanau, Cho, and Bengio}{Bahdanau
  et~al\mbox{.}}{2015}]%
        {Bahdanau:2015:NMT}
\bibfield{author}{\bibinfo{person}{Dzmitry Bahdanau},
  \bibinfo{person}{Kyunghyun Cho}, {and} \bibinfo{person}{Yoshua Bengio}.}
  \bibinfo{year}{2015}\natexlab{}.
\newblock \showarticletitle{Neural Machine Translation by Jointly Learning to
  Align and Translate}. In \bibinfo{booktitle}{\emph{Proc. of ICLR '15}}.
\newblock


\bibitem[\protect\citeauthoryear{Bao, Duan, Yan, Zhou, and Zhao}{Bao
  et~al\mbox{.}}{2016}]%
        {Bao:2016:CBQ}
\bibfield{author}{\bibinfo{person}{Junwei Bao}, \bibinfo{person}{Nan Duan},
  \bibinfo{person}{Zhao Yan}, \bibinfo{person}{Ming Zhou}, {and}
  \bibinfo{person}{Tiejun Zhao}.} \bibinfo{year}{2016}\natexlab{}.
\newblock \showarticletitle{Constraint-Based Question Answering with Knowledge
  Graph}. In \bibinfo{booktitle}{\emph{Proc. of COLING '16}}.
  \bibinfo{pages}{2503--2514}.
\newblock


\bibitem[\protect\citeauthoryear{Berant, Chou, Frostig, and Liang}{Berant
  et~al\mbox{.}}{2013}]%
        {Berant:2013:SPF}
\bibfield{author}{\bibinfo{person}{Jonathan Berant}, \bibinfo{person}{Andrew
  Chou}, \bibinfo{person}{Roy Frostig}, {and} \bibinfo{person}{Percy Liang}.}
  \bibinfo{year}{2013}\natexlab{}.
\newblock \showarticletitle{Semantic Parsing on {F}reebase from Question-Answer
  Pairs}. In \bibinfo{booktitle}{\emph{Proc. of EMNLP '13}}.
  \bibinfo{pages}{1533--1544}.
\newblock


\bibitem[\protect\citeauthoryear{Bordes, Usunier, Chopra, and Weston}{Bordes
  et~al\mbox{.}}{2015}]%
        {Bordes:2015:LSQ}
\bibfield{author}{\bibinfo{person}{Antoine Bordes}, \bibinfo{person}{Nicolas
  Usunier}, \bibinfo{person}{Sumit Chopra}, {and} \bibinfo{person}{Jason
  Weston}.} \bibinfo{year}{2015}\natexlab{}.
\newblock \showarticletitle{Large-scale Simple Question Answering with Memory
  Networks}.
\newblock  (\bibinfo{year}{2015}).
\newblock
\showeprint[arxiv]{1506.02075}


\bibitem[\protect\citeauthoryear{Brown, Della~Pietra, Della~Pietra, Lai, and
  Mercer}{Brown et~al\mbox{.}}{1992}]%
        {Brown:1992:EUB}
\bibfield{author}{\bibinfo{person}{Peter~F. Brown}, \bibinfo{person}{Stephen~A.
  Della~Pietra}, \bibinfo{person}{Vincent~J. Della~Pietra},
  \bibinfo{person}{Jennifer~C. Lai}, {and} \bibinfo{person}{Robert~L. Mercer}.}
  \bibinfo{year}{1992}\natexlab{}.
\newblock \showarticletitle{An Estimate of an Upper Bound for the Entropy of
  {E}nglish}.
\newblock \bibinfo{journal}{\emph{Computational Linguistics}}
  \bibinfo{volume}{18}, \bibinfo{number}{1} (\bibinfo{year}{1992}),
  \bibinfo{pages}{31--40}.
\newblock


\bibitem[\protect\citeauthoryear{Cai and Yates}{Cai and Yates}{2013}]%
        {Cai:2013:LSP}
\bibfield{author}{\bibinfo{person}{Qingqing Cai} {and}
  \bibinfo{person}{Alexander Yates}.} \bibinfo{year}{2013}\natexlab{}.
\newblock \showarticletitle{Large-scale Semantic Parsing via Schema Matching
  and Lexicon Extension}. In \bibinfo{booktitle}{\emph{Proc. of ACL '13}}.
  \bibinfo{pages}{423--433}.
\newblock


\bibitem[\protect\citeauthoryear{Chakraborty, Lukovnikov, Maheshwari, Trivedi,
  Lehmann, and Fischer}{Chakraborty et~al\mbox{.}}{2019}]%
        {Chakraborty:2019:INN}
\bibfield{author}{\bibinfo{person}{Nilesh Chakraborty}, \bibinfo{person}{Denis
  Lukovnikov}, \bibinfo{person}{Gaurav Maheshwari}, \bibinfo{person}{Priyansh
  Trivedi}, \bibinfo{person}{Jens Lehmann}, {and} \bibinfo{person}{Asja
  Fischer}.} \bibinfo{year}{2019}\natexlab{}.
\newblock \showarticletitle{Introduction to Neural Network based Approaches for
  Question Answering over Knowledge Graphs}.
\newblock  (\bibinfo{year}{2019}).
\newblock
\showeprint[arxiv]{1907.09361}


\bibitem[\protect\citeauthoryear{Dao, Gu, Ratner, Smith, De~Sa, and Re}{Dao
  et~al\mbox{.}}{2019}]%
        {Dao:2019:KTM}
\bibfield{author}{\bibinfo{person}{Tri Dao}, \bibinfo{person}{Albert Gu},
  \bibinfo{person}{Alexander Ratner}, \bibinfo{person}{Virginia Smith},
  \bibinfo{person}{Chris De~Sa}, {and} \bibinfo{person}{Christopher Re}.}
  \bibinfo{year}{2019}\natexlab{}.
\newblock \showarticletitle{A Kernel Theory of Modern Data Augmentation}. In
  \bibinfo{booktitle}{\emph{Proc. of ICML '19}}
  \emph{(\bibinfo{series}{PMLR})}, Vol.~\bibinfo{volume}{97}.
  \bibinfo{pages}{1528--1537}.
\newblock


\bibitem[\protect\citeauthoryear{Devlin, Chang, Lee, and Toutanova}{Devlin
  et~al\mbox{.}}{2019}]%
        {Devlin:2019:BERT}
\bibfield{author}{\bibinfo{person}{Jacob Devlin}, \bibinfo{person}{Ming-Wei
  Chang}, \bibinfo{person}{Kenton Lee}, {and} \bibinfo{person}{Kristina
  Toutanova}.} \bibinfo{year}{2019}\natexlab{}.
\newblock \showarticletitle{{BERT}: Pre-training of Deep Bidirectional
  Transformers for Language Understanding}. In \bibinfo{booktitle}{\emph{Proc.
  of NAACL '19}}. \bibinfo{pages}{4171--4186}.
\newblock


\bibitem[\protect\citeauthoryear{Diefenbach, L{\'o}pez, Singh, and
  Maret}{Diefenbach et~al\mbox{.}}{2017}]%
        {Diefenbach:2017:CTQ}
\bibfield{author}{\bibinfo{person}{Dennis Diefenbach}, \bibinfo{person}{Vanessa
  L{\'o}pez}, \bibinfo{person}{Kamal~Deep Singh}, {and} \bibinfo{person}{Pierre
  Maret}.} \bibinfo{year}{2017}\natexlab{}.
\newblock \showarticletitle{Core techniques of question answering systems over
  knowledge bases: a survey}.
\newblock \bibinfo{journal}{\emph{Knowledge and Information Systems}}
  \bibinfo{volume}{55} (\bibinfo{year}{2017}), \bibinfo{pages}{529--569}.
\newblock


\bibitem[\protect\citeauthoryear{Ding and Balog}{Ding and Balog}{2018}]%
        {Ding:2018:GSD}
\bibfield{author}{\bibinfo{person}{Heng Ding} {and} \bibinfo{person}{Krisztian
  Balog}.} \bibinfo{year}{2018}\natexlab{}.
\newblock \showarticletitle{Generating Synthetic Data for Neural
  Keyword-to-Question Models}. In \bibinfo{booktitle}{\emph{Proc. of ICTIR
  '18}}. \bibinfo{pages}{51--58}.
\newblock


\bibitem[\protect\citeauthoryear{Dong and Lapata}{Dong and Lapata}{2016}]%
        {DongLapata:2016:LLF}
\bibfield{author}{\bibinfo{person}{Li Dong} {and} \bibinfo{person}{Mirella
  Lapata}.} \bibinfo{year}{2016}\natexlab{}.
\newblock \showarticletitle{Language to Logical Form with Neural Attention}. In
  \bibinfo{booktitle}{\emph{Proc. of ACL '16}}. \bibinfo{pages}{33--43}.
\newblock


\bibitem[\protect\citeauthoryear{Dubey, Banerjee, Abdelkawi, and Lehmann}{Dubey
  et~al\mbox{.}}{2019}]%
        {Dubey:2019:LCQ}
\bibfield{author}{\bibinfo{person}{Mohnish Dubey}, \bibinfo{person}{Debayan
  Banerjee}, \bibinfo{person}{Abdelrahman Abdelkawi}, {and}
  \bibinfo{person}{Jens Lehmann}.} \bibinfo{year}{2019}\natexlab{}.
\newblock \showarticletitle{LC-QuAD 2.0: A Large Dataset for Complex Question
  Answering over Wikidata and DBpedia}. In \bibinfo{booktitle}{\emph{Proc. of
  ISWC '19}}, \bibfield{editor}{\bibinfo{person}{Chiara Ghidini},
  \bibinfo{person}{Olaf Hartig}, \bibinfo{person}{Maria Maleshkova},
  \bibinfo{person}{Vojt{\v{e}}ch Sv{\'a}tek}, \bibinfo{person}{Isabel Cruz},
  \bibinfo{person}{Aidan Hogan}, \bibinfo{person}{Jie Song},
  \bibinfo{person}{Maxime Lefran{\c{c}}ois}, {and} \bibinfo{person}{Fabien
  Gandon}} (Eds.). \bibinfo{pages}{69--78}.
\newblock


\bibitem[\protect\citeauthoryear{Gehring, Auli, Grangier, Yarats, and
  Dauphin}{Gehring et~al\mbox{.}}{2017}]%
        {Gehring:2017:CSS}
\bibfield{author}{\bibinfo{person}{Jonas Gehring}, \bibinfo{person}{Michael
  Auli}, \bibinfo{person}{David Grangier}, \bibinfo{person}{Denis Yarats},
  {and} \bibinfo{person}{Yann~N. Dauphin}.} \bibinfo{year}{2017}\natexlab{}.
\newblock \showarticletitle{Convolutional Sequence to Sequence Learning}. In
  \bibinfo{booktitle}{\emph{Proc. of ICML '17}}. \bibinfo{pages}{1243--1252}.
\newblock


\bibitem[\protect\citeauthoryear{Golub, Huang, He, and Deng}{Golub
  et~al\mbox{.}}{2017}]%
        {Golub:2017:TSS}
\bibfield{author}{\bibinfo{person}{David Golub}, \bibinfo{person}{Po-Sen
  Huang}, \bibinfo{person}{Xiaodong He}, {and} \bibinfo{person}{Li Deng}.}
  \bibinfo{year}{2017}\natexlab{}.
\newblock \showarticletitle{Two-Stage Synthesis Networks for Transfer Learning
  in Machine Comprehension}. In \bibinfo{booktitle}{\emph{Proc. of EMNLP '17}}.
  \bibinfo{pages}{835--844}.
\newblock


\bibitem[\protect\citeauthoryear{Goodfellow, Pouget-Abadie, Mirza, Xu,
  Warde-Farley, Ozair, Courville, and Bengio}{Goodfellow et~al\mbox{.}}{2014}]%
        {Goodfellow:2014:GAN}
\bibfield{author}{\bibinfo{person}{Ian Goodfellow}, \bibinfo{person}{Jean
  Pouget-Abadie}, \bibinfo{person}{Mehdi Mirza}, \bibinfo{person}{Bing Xu},
  \bibinfo{person}{David Warde-Farley}, \bibinfo{person}{Sherjil Ozair},
  \bibinfo{person}{Aaron Courville}, {and} \bibinfo{person}{Yoshua Bengio}.}
  \bibinfo{year}{2014}\natexlab{}.
\newblock \showarticletitle{Generative Adversarial Nets}.
\newblock In \bibinfo{booktitle}{\emph{Advances in Neural Information
  Processing Systems 27}}, \bibfield{editor}{\bibinfo{person}{Z.~Ghahramani},
  \bibinfo{person}{M.~Welling}, \bibinfo{person}{C.~Cortes},
  \bibinfo{person}{N.~D. Lawrence}, {and} \bibinfo{person}{K.~Q. Weinberger}}
  (Eds.). \bibinfo{pages}{2672--2680}.
\newblock


\bibitem[\protect\citeauthoryear{Hartmann, Marx, and Soru}{Hartmann
  et~al\mbox{.}}{2018}]%
        {Hartmann:2018:GLD}
\bibfield{author}{\bibinfo{person}{Ann-Kathrin Hartmann},
  \bibinfo{person}{Edgard Marx}, {and} \bibinfo{person}{Tommaso Soru}.}
  \bibinfo{year}{2018}\natexlab{}.
\newblock \showarticletitle{Generating a Large Dataset for Neural Question
  Answering over the {DB}pedia Knowledge Base}.
\newblock  (\bibinfo{year}{2018}).
\newblock


\bibitem[\protect\citeauthoryear{Jia and Liang}{Jia and Liang}{2016}]%
        {JiaLiang:2016:DRN}
\bibfield{author}{\bibinfo{person}{Robin Jia} {and} \bibinfo{person}{Percy
  Liang}.} \bibinfo{year}{2016}\natexlab{}.
\newblock \showarticletitle{Data Recombination for Neural Semantic Parsing}. In
  \bibinfo{booktitle}{\emph{Proc. of ACL '16}}. \bibinfo{pages}{12--22}.
\newblock


\bibitem[\protect\citeauthoryear{Kingma and Welling}{Kingma and
  Welling}{2014}]%
        {Diederik:2014:AEV}
\bibfield{author}{\bibinfo{person}{Diederik~P. Kingma} {and}
  \bibinfo{person}{Max Welling}.} \bibinfo{year}{2014}\natexlab{}.
\newblock \showarticletitle{Auto-Encoding Variational Bayes}. In
  \bibinfo{booktitle}{\emph{Proc. of ICLR '14}}.
\newblock


\bibitem[\protect\citeauthoryear{Krishnan, Moffat, Zobel, and
  Billerbeck}{Krishnan et~al\mbox{.}}{2020}]%
        {Krishnan:2020:GSQ}
\bibfield{author}{\bibinfo{person}{Unni Krishnan}, \bibinfo{person}{Alistair
  Moffat}, \bibinfo{person}{Justin Zobel}, {and} \bibinfo{person}{Bodo
  Billerbeck}.} \bibinfo{year}{2020}\natexlab{}.
\newblock \showarticletitle{Generation of Synthetic Query Auto Completion
  Logs}. In \bibinfo{booktitle}{\emph{Proc. of ECIR '20}},
  \bibfield{editor}{\bibinfo{person}{Joemon~M. Jose}, \bibinfo{person}{Emine
  Yilmaz}, \bibinfo{person}{Jo{\~a}o Magalh{\~a}es}, \bibinfo{person}{Pablo
  Castells}, \bibinfo{person}{Nicola Ferro}, \bibinfo{person}{M{\'a}rio~J.
  Silva}, {and} \bibinfo{person}{Fl{\'a}vio Martins}} (Eds.).
  \bibinfo{pages}{621--635}.
\newblock


\bibitem[\protect\citeauthoryear{Liang}{Liang}{2013}]%
        {Liang:2013:LDC}
\bibfield{author}{\bibinfo{person}{P. Liang}.} \bibinfo{year}{2013}\natexlab{}.
\newblock \showarticletitle{Lambda Dependency-Based Compositional Semantics}.
\newblock  (\bibinfo{year}{2013}).
\newblock
\showeprint[arxiv]{1309.4408}


\bibitem[\protect\citeauthoryear{Linjordet and Balog}{Linjordet and
  Balog}{2019}]%
        {Linjordet:2019:ITD}
\bibfield{author}{\bibinfo{person}{Trond Linjordet} {and}
  \bibinfo{person}{Krisztian Balog}.} \bibinfo{year}{2019}\natexlab{}.
\newblock \showarticletitle{Impact of Training Dataset Size on Neural Answer
  Selection Models}. In \bibinfo{booktitle}{\emph{Proc. of ECIR '19}}.
  \bibinfo{pages}{828--835}.
\newblock


\bibitem[\protect\citeauthoryear{Luong, Pham, and Manning}{Luong
  et~al\mbox{.}}{2015}]%
        {Luong:2015:EAA}
\bibfield{author}{\bibinfo{person}{Thang Luong}, \bibinfo{person}{Hieu Pham},
  {and} \bibinfo{person}{Christopher~D. Manning}.}
  \bibinfo{year}{2015}\natexlab{}.
\newblock \showarticletitle{Effective Approaches to Attention-based Neural
  Machine Translation}. In \bibinfo{booktitle}{\emph{Proc. of EMNLP '15}}.
  \bibinfo{pages}{1412--1421}.
\newblock


\bibitem[\protect\citeauthoryear{Nikolenko}{Nikolenko}{2019}]%
        {Nikolenko:2019:SDD}
\bibfield{author}{\bibinfo{person}{Sergey~I. Nikolenko}.}
  \bibinfo{year}{2019}\natexlab{}.
\newblock \bibinfo{title}{Synthetic Data for Deep Learning}.
\newblock
\newblock
\showeprint[arxiv]{1909.11512}


\bibitem[\protect\citeauthoryear{Papineni, Roukos, Ward, and Zhu}{Papineni
  et~al\mbox{.}}{2002}]%
        {Papineni:2002:BLEU}
\bibfield{author}{\bibinfo{person}{Kishore Papineni}, \bibinfo{person}{Salim
  Roukos}, \bibinfo{person}{Todd Ward}, {and} \bibinfo{person}{Wei-Jing Zhu}.}
  \bibinfo{year}{2002}\natexlab{}.
\newblock \showarticletitle{{B}leu: a Method for Automatic Evaluation of
  Machine Translation}. In \bibinfo{booktitle}{\emph{Proc. of ACL '02}}.
  \bibinfo{pages}{311--318}.
\newblock


\bibitem[\protect\citeauthoryear{Reddy, Lapata, and Steedman}{Reddy
  et~al\mbox{.}}{2014}]%
        {Reddy:2014:LSS}
\bibfield{author}{\bibinfo{person}{Siva Reddy}, \bibinfo{person}{Mirella
  Lapata}, {and} \bibinfo{person}{Mark Steedman}.}
  \bibinfo{year}{2014}\natexlab{}.
\newblock \showarticletitle{Large-scale Semantic Parsing without
  Question-Answer Pairs}.
\newblock  (\bibinfo{year}{2014}).
\newblock


\bibitem[\protect\citeauthoryear{Serban, Garc{\'\i}a-Dur{\'a}n, Gulcehre, Ahn,
  Chandar, Courville, and Bengio}{Serban et~al\mbox{.}}{2016}]%
        {Serban:2016:GFQ}
\bibfield{author}{\bibinfo{person}{Iulian~Vlad Serban},
  \bibinfo{person}{Alberto Garc{\'\i}a-Dur{\'a}n}, \bibinfo{person}{Caglar
  Gulcehre}, \bibinfo{person}{Sungjin Ahn}, \bibinfo{person}{Sarath Chandar},
  \bibinfo{person}{Aaron Courville}, {and} \bibinfo{person}{Yoshua Bengio}.}
  \bibinfo{year}{2016}\natexlab{}.
\newblock \showarticletitle{Generating Factoid Questions With Recurrent Neural
  Networks: The 30{M} Factoid Question-Answer Corpus}. In
  \bibinfo{booktitle}{\emph{Proc. of ACL '16}}. \bibinfo{pages}{588--598}.
\newblock


\bibitem[\protect\citeauthoryear{Shorten and Khoshgoftaar}{Shorten and
  Khoshgoftaar}{2019}]%
        {Shorten:2019:SID}
\bibfield{author}{\bibinfo{person}{Connor Shorten} {and}
  \bibinfo{person}{Taghi~M. Khoshgoftaar}.} \bibinfo{year}{2019}\natexlab{}.
\newblock \showarticletitle{A Survey on Image Data Augmentation for Deep
  Learning}.
\newblock \bibinfo{journal}{\emph{Journal of Big Data}}  \bibinfo{volume}{6},
  Article \bibinfo{articleno}{60} (\bibinfo{year}{2019}).
\newblock


\bibitem[\protect\citeauthoryear{Soru, Marx, Moussallem, Publio, Valdestilhas,
  Esteves, and Neto}{Soru et~al\mbox{.}}{2017}]%
        {Soru:2017:SFL}
\bibfield{author}{\bibinfo{person}{Tommaso Soru}, \bibinfo{person}{Edgard
  Marx}, \bibinfo{person}{Diego Moussallem}, \bibinfo{person}{Gustavo Publio},
  \bibinfo{person}{Andr\'e Valdestilhas}, \bibinfo{person}{Diego Esteves},
  {and} \bibinfo{person}{Ciro~Baron Neto}.} \bibinfo{year}{2017}\natexlab{}.
\newblock \showarticletitle{{SPARQL} as a Foreign Language}. In
  \bibinfo{booktitle}{\emph{Proc. of SEMANTiCS '17 - Posters and Demos}}.
\newblock


\bibitem[\protect\citeauthoryear{Soru, Marx, Valdestilhas, Esteves, Moussallem,
  and Publio}{Soru et~al\mbox{.}}{2018}]%
        {Soru:2018:NMT}
\bibfield{author}{\bibinfo{person}{Tommaso Soru}, \bibinfo{person}{Edgard
  Marx}, \bibinfo{person}{Andr\'e Valdestilhas}, \bibinfo{person}{Diego
  Esteves}, \bibinfo{person}{Diego Moussallem}, {and} \bibinfo{person}{Gustavo
  Publio}.} \bibinfo{year}{2018}\natexlab{}.
\newblock \showarticletitle{Neural Machine Translation for Query Construction
  and Composition}. In \bibinfo{booktitle}{\emph{Proc. of ICML '18 - Workshop
  on Neural Abstract Machines \& Program Induction (NAMPI v2)}}.
\newblock


\bibitem[\protect\citeauthoryear{Su, Sun, Sadler, Srivatsa, G{\"u}r, Yan, and
  Yan}{Su et~al\mbox{.}}{2016}]%
        {Su-2016-GCQ}
\bibfield{author}{\bibinfo{person}{Yu Su}, \bibinfo{person}{Huan Sun},
  \bibinfo{person}{Brian Sadler}, \bibinfo{person}{Mudhakar Srivatsa},
  \bibinfo{person}{Izzeddin G{\"u}r}, \bibinfo{person}{Zenghui Yan}, {and}
  \bibinfo{person}{Xifeng Yan}.} \bibinfo{year}{2016}\natexlab{}.
\newblock \showarticletitle{On Generating Characteristic-rich Question Sets for
  {QA} Evaluation}. In \bibinfo{booktitle}{\emph{Proc. of EMNLP '16}}.
  \bibinfo{pages}{562--572}.
\newblock


\bibitem[\protect\citeauthoryear{Talmor and Berant}{Talmor and Berant}{2018}]%
        {Talmor:2018:WKB}
\bibfield{author}{\bibinfo{person}{Alon Talmor} {and} \bibinfo{person}{Jonathan
  Berant}.} \bibinfo{year}{2018}\natexlab{}.
\newblock \showarticletitle{The Web as a Knowledge-Base for Answering Complex
  Questions}. In \bibinfo{booktitle}{\emph{Proc. of NAACL '18}}.
  \bibinfo{pages}{641--651}.
\newblock


\bibitem[\protect\citeauthoryear{Trivedi, Maheshwari, Dubey, and
  Lehmann}{Trivedi et~al\mbox{.}}{2017}]%
        {Trivedi:2017:LCQ}
\bibfield{author}{\bibinfo{person}{Priyansh Trivedi}, \bibinfo{person}{Gaurav
  Maheshwari}, \bibinfo{person}{Mohnish Dubey}, {and} \bibinfo{person}{Jens
  Lehmann}.} \bibinfo{year}{2017}\natexlab{}.
\newblock \showarticletitle{LC-QuAD: A Corpus for Complex Question Answering
  over Knowledge Graphs}. In \bibinfo{booktitle}{\emph{Proc. of ISWC '17}},
  \bibfield{editor}{\bibinfo{person}{Claudia d'Amato}, \bibinfo{person}{Miriam
  Fernandez}, \bibinfo{person}{Valentina Tamma}, \bibinfo{person}{Freddy
  Lecue}, \bibinfo{person}{Philippe Cudr{\'e}-Mauroux}, \bibinfo{person}{Juan
  Sequeda}, \bibinfo{person}{Christoph Lange}, {and} \bibinfo{person}{Jeff
  Heflin}} (Eds.). \bibinfo{pages}{210--218}.
\newblock


\bibitem[\protect\citeauthoryear{Unger, Forascu, L{\'o}pez, Ngomo, Cabrio,
  Cimiano, and Walter}{Unger et~al\mbox{.}}{2014}]%
        {Unger:2014:QAL}
\bibfield{author}{\bibinfo{person}{Christina Unger}, \bibinfo{person}{Corina
  Forascu}, \bibinfo{person}{Vanessa L{\'o}pez},
  \bibinfo{person}{Axel-Cyrille~Ngonga Ngomo}, \bibinfo{person}{Elena Cabrio},
  \bibinfo{person}{Philipp Cimiano}, {and} \bibinfo{person}{Sebastian Walter}.}
  \bibinfo{year}{2014}\natexlab{}.
\newblock \showarticletitle{Question Answering over Linked Data (QALD-5)}. In
  \bibinfo{booktitle}{\emph{Proc. of CLEF '14}}.
\newblock


\bibitem[\protect\citeauthoryear{Usbeck, Ngomo, Haarmann, Krithara, R\"{o}der,
  and Napolitano}{Usbeck et~al\mbox{.}}{2017}]%
        {Usbeck:2017:7OC}
\bibfield{author}{\bibinfo{person}{Ricardo Usbeck},
  \bibinfo{person}{Axel-Cyrille~Ngonga Ngomo}, \bibinfo{person}{Bastian
  Haarmann}, \bibinfo{person}{Anastasia Krithara}, \bibinfo{person}{Michael
  R\"{o}der}, {and} \bibinfo{person}{Giulio Napolitano}.}
  \bibinfo{year}{2017}\natexlab{}.
\newblock \showarticletitle{7th Open Challenge on Question Answering over
  Linked Data (QALD-7)}. In \bibinfo{booktitle}{\emph{Semantic Web Challenges}}
  \emph{(\bibinfo{series}{SemWebEval '17})}, Vol.~\bibinfo{volume}{769}.
\newblock


\bibitem[\protect\citeauthoryear{Vakulenko, Fernandez~Garcia, Polleres,
  de~Rijke, and Cochez}{Vakulenko et~al\mbox{.}}{2019}]%
        {Vakulenko:2019:MPC}
\bibfield{author}{\bibinfo{person}{Svitlana Vakulenko},
  \bibinfo{person}{Javier~David Fernandez~Garcia}, \bibinfo{person}{Axel
  Polleres}, \bibinfo{person}{Maarten de Rijke}, {and} \bibinfo{person}{Michael
  Cochez}.} \bibinfo{year}{2019}\natexlab{}.
\newblock \showarticletitle{Message Passing for Complex Question Answering over
  Knowledge Graphs}. In \bibinfo{booktitle}{\emph{Proc. of CIKM '19}}.
  \bibinfo{pages}{1431--1440}.
\newblock


\bibitem[\protect\citeauthoryear{Vaswani, Shazeer, Parmar, Uszkoreit, Jones,
  Gomez, Kaiser, and Polosukhin}{Vaswani et~al\mbox{.}}{2017}]%
        {Vaswani:2017:AIA}
\bibfield{author}{\bibinfo{person}{Ashish Vaswani}, \bibinfo{person}{Noam
  Shazeer}, \bibinfo{person}{Niki Parmar}, \bibinfo{person}{Jakob Uszkoreit},
  \bibinfo{person}{Llion Jones}, \bibinfo{person}{Aidan~N Gomez},
  \bibinfo{person}{Lukasz Kaiser}, {and} \bibinfo{person}{Illia Polosukhin}.}
  \bibinfo{year}{2017}\natexlab{}.
\newblock \showarticletitle{Attention is All you Need}.
\newblock In \bibinfo{booktitle}{\emph{Advances in Neural Information
  Processing Systems 30}}, \bibfield{editor}{\bibinfo{person}{I.~Guyon},
  \bibinfo{person}{U.~V. Luxburg}, \bibinfo{person}{S.~Bengio},
  \bibinfo{person}{H.~Wallach}, \bibinfo{person}{R.~Fergus},
  \bibinfo{person}{S.~Vishwanathan}, {and} \bibinfo{person}{R.~Garnett}}
  (Eds.). \bibinfo{pages}{5998--6008}.
\newblock


\bibitem[\protect\citeauthoryear{Wu, Schuster, Chen, Le, Norouzi, Macherey,
  Krikun, Cao, Gao, Macherey, Klingner, Shah, Johnson, Liu, Łukasz Kaiser,
  Gouws, Kato, Kudo, Kazawa, Stevens, Kurian, Patil, Wang, Young, Smith, Riesa,
  Rudnick, Vinyals, Corrado, Hughes, and Dean}{Wu et~al\mbox{.}}{2016}]%
        {Wu:2016:GNM}
\bibfield{author}{\bibinfo{person}{Yonghui Wu}, \bibinfo{person}{Mike
  Schuster}, \bibinfo{person}{Zhifeng Chen}, \bibinfo{person}{Quoc~V. Le},
  \bibinfo{person}{Mohammad Norouzi}, \bibinfo{person}{Wolfgang Macherey},
  \bibinfo{person}{Maxim Krikun}, \bibinfo{person}{Yuan Cao},
  \bibinfo{person}{Qin Gao}, \bibinfo{person}{Klaus Macherey},
  \bibinfo{person}{Jeff Klingner}, \bibinfo{person}{Apurva Shah},
  \bibinfo{person}{Melvin Johnson}, \bibinfo{person}{Xiaobing Liu},
  \bibinfo{person}{Łukasz Kaiser}, \bibinfo{person}{Stephan Gouws},
  \bibinfo{person}{Yoshikiyo Kato}, \bibinfo{person}{Taku Kudo},
  \bibinfo{person}{Hideto Kazawa}, \bibinfo{person}{Keith Stevens},
  \bibinfo{person}{George Kurian}, \bibinfo{person}{Nishant Patil},
  \bibinfo{person}{Wei Wang}, \bibinfo{person}{Cliff Young},
  \bibinfo{person}{Jason Smith}, \bibinfo{person}{Jason Riesa},
  \bibinfo{person}{Alex Rudnick}, \bibinfo{person}{Oriol Vinyals},
  \bibinfo{person}{Greg Corrado}, \bibinfo{person}{Macduff Hughes}, {and}
  \bibinfo{person}{Jeffrey Dean}.} \bibinfo{year}{2016}\natexlab{}.
\newblock \showarticletitle{Google's Neural Machine Translation System:
  Bridging the Gap between Human and Machine Translation}.
\newblock  (\bibinfo{year}{2016}).
\newblock
\showeprint[arxiv]{1609.08144}


\bibitem[\protect\citeauthoryear{Yang, Huo, Shen, Cheng, Wang, Wang, and
  Carin}{Yang et~al\mbox{.}}{2019}]%
        {Yang:2019:EEG}
\bibfield{author}{\bibinfo{person}{Qian Yang}, \bibinfo{person}{Zhouyuan Huo},
  \bibinfo{person}{Dinghan Shen}, \bibinfo{person}{Yong Cheng},
  \bibinfo{person}{Wenlin Wang}, \bibinfo{person}{Guoyin Wang}, {and}
  \bibinfo{person}{Lawrence Carin}.} \bibinfo{year}{2019}\natexlab{}.
\newblock \showarticletitle{An End-to-End Generative Architecture for
  Paraphrase Generation}. In \bibinfo{booktitle}{\emph{Proc. of EMNLP-IJCNLP
  '19}}. \bibinfo{pages}{3132--3142}.
\newblock


\bibitem[\protect\citeauthoryear{Yih, Chang, He, and Gao}{Yih
  et~al\mbox{.}}{2015}]%
        {Yih:2015:SPS}
\bibfield{author}{\bibinfo{person}{Scott Wen-tau Yih},
  \bibinfo{person}{Ming-Wei Chang}, \bibinfo{person}{Xiaodong He}, {and}
  \bibinfo{person}{Jianfeng Gao}.} \bibinfo{year}{2015}\natexlab{}.
\newblock \showarticletitle{Semantic Parsing via Staged Query Graph Generation:
  Question Answering with Knowledge Base}. In \bibinfo{booktitle}{\emph{Proc.
  of AACL-IJCNLP '15}}.
\newblock


\bibitem[\protect\citeauthoryear{Yih, Richardson, Meek, Chang, and Suh}{Yih
  et~al\mbox{.}}{[n.d.]}]%
        {Yih:2016:VSP}
\bibfield{author}{\bibinfo{person}{Wen-tau Yih}, \bibinfo{person}{Matthew
  Richardson}, \bibinfo{person}{Chris Meek}, \bibinfo{person}{Ming-Wei Chang},
  {and} \bibinfo{person}{Jina Suh}.} \bibinfo{year}{[n.d.]}\natexlab{}.
\newblock \showarticletitle{The Value of Semantic Parse Labeling for Knowledge
  Base Question Answering}. In \bibinfo{booktitle}{\emph{Proc. of ACL '16}}.
\newblock


\bibitem[\protect\citeauthoryear{Yin, Gromann, and Rudolph}{Yin
  et~al\mbox{.}}{2019}]%
        {Yin:2019:NMT}
\bibfield{author}{\bibinfo{person}{Xiaoyu Yin}, \bibinfo{person}{Dagmar
  Gromann}, {and} \bibinfo{person}{Sebastian Rudolph}.}
  \bibinfo{year}{2019}\natexlab{}.
\newblock \showarticletitle{Neural Machine Translating from Natural Language to
  SPARQL}.
\newblock  (\bibinfo{year}{2019}).
\newblock


\bibitem[\protect\citeauthoryear{Zhang and Bansal}{Zhang and Bansal}{2019}]%
        {Zhang:2019:ASD}
\bibfield{author}{\bibinfo{person}{Shiyue Zhang} {and} \bibinfo{person}{Mohit
  Bansal}.} \bibinfo{year}{2019}\natexlab{}.
\newblock \showarticletitle{Addressing Semantic Drift in Question Generation
  for Semi-Supervised Question Answering}. In \bibinfo{booktitle}{\emph{Proc.
  of EMNLP-IJCNLP '19}}. \bibinfo{pages}{2495--2509}.
\newblock


\end{thebibliography}
